**Anisotropic field-induced changes in the superconducting order parameter of UTe$_2$**

**Sangyun Lee[1,¶], Andrew J. Woods[1,¶], P. F. S. Rosa[1], S. M. Thomas,**

**E. D. Bauer[1], Shi-Zeng Lin[1], and R. Movshovich[1,*]**

[1]Los Alamos National Laboratory, Los Alamos, New Mexico 87545, USA

[¶] These authors contributed equally to this work.

Corresponding author: R. M. (roman@lanl.gov)

**Abstract**

UTe$_2$ is a newly discovered unconventional superconductor, where electron Cooper pairs combine into a spin-triplet ground state. Here we report the specific heat $C(H,T)$ of a high-quality single crystal of UTe$_2$ with a single specific heat anomaly at the superconducting transition temperature $T_c \approx 2$ K and a small zero-field residual Sommerfeld coefficient $\gamma_0 = C/T$ $(T=0) \approx 10$ mJ/mol-K$^2$. We applied magnetic field up to 12 T along the three principal crystallographic axes of UTe$_2$ to probe the nature of the superconducting state. The evolution of the residual Sommerfeld coefficient as a function of magnetic field, $\gamma_0$ $(H)$, is highly anisotropic and reveals distinct regions. In magnetic field up to 4 T applied along $a$, $b$, and $c$ axes, we find $\gamma_0 \approx \alpha_i\sqrt{H}$, with $i = a, b, c$, as expected for an unconventional superconductor with nodes (zeros) of the superconducting order parameter on the Fermi surface. A pronounced kink in $\gamma_0(H)$, however, is observed at roughly 4 T for field applied along both $a$ and $b$ axes, whereas a smooth change from square-root to linear behaviour is observed at 4 T for $H$||$c$. These results strongly indicate that a zero-field ground state is stable up to 4 T and undergoes

a field-induced evolution above 4 T. $\alpha_c > \alpha_a > \alpha_b$, indicating that the nodes in the low-field state are predominantly located in the vicinity of the $a - b$ plane. The modification of the order parameter is strongest when field is applied in the $a - b$ plane, which causes nodes to move away from the direction of the applied field. Both $\boldsymbol{d}_{B2u}+i\varepsilon\boldsymbol{d}_{B1u}$ and $\boldsymbol{d}_{B2u}+i\varepsilon\boldsymbol{d}_{Au}$ two-component order parameters can account for our observations, with $\boldsymbol{d}_{B2u}+i\varepsilon\boldsymbol{d}_{B1u}$ a more likely candidate. In either scenario, our measurements indicate that $B_{2u}$ is the primary superconducting order parameter in UTe$_2$.

**Introduction**

The experimental search for odd-parity superconductivity has intensified in recent years due to the potential for this state to host Majorana modes, which could be used to realize fault-tolerant quantum computers. [1-4]. A Cooper pair in an odd-parity superconductor has spin 1 and odd orbital parity, which can be robust against applied magnetic field compared to its even-parity counterpart. In the recently-discovered superconductor UTe$_2$, the upper critical fields ($H_{c2}$) surpass the Pauli limit for weakly coupled superconductors without spin-orbit coupling (see, e.g. ref. [5]), and a reentrant superconducting state is observed for field applied along the $b$ axis of this orthorhombic system [6].

A prototypical odd-parity superconducting ground state was discovered experimentally and described theoretically in superfluid $^3$He [7,8]. A number of solid-state compounds have been suggested to host odd-parity superconducting ground states. One intensely investigated example is Sr$_2$RuO$_4$, whose superconducting state was recently shown by nuclear magnetic resonance measurements to be even parity [9,10]. A fruitful avenue to odd-parity superconductivity may be found in uranium-based ferromagnetic (FM) compounds, wherein superconductivity is realized near a ferromagnetic quantum critical point [11]. Several U-based superconductivity, such as $UGe_2$, UCoGe, and URhGe, appear to fit this scenario well, with superconductivity developing within the FM ordered state [12-16]. We note, however, that odd-parity superconductor $UPt_3$ does not show signs of ferromagnetism and therefore may not fit this scenario. [17]

Based on several investigations, it was proposed that $UTe_2$ is an end member of the family of FM superconductors without a FM ground state [18]. µSR studies of $UTe_2$ showed evidence of FM fluctuations, although no long-range magnetic order has been observed down to 25 mK [19]. Nuclear magnetic resonance (NMR) measurements revealed a small decrease in Knight shift below $T_c$, which is smaller than that expected for a spin-singlet state [20]. Combined with the observation that $H_{c2}$ greatly exceeds the Pauli limit [21-23], these results indicate that $UTe_2$ is a prime candidate for an odd-parity superconductor. Initial studies of the specific heat, thermal conductivity, and effective penetration depth on lower $T_c$ samples indicated nodal superconductivity with either line or point nodes in the *a-b* plane [24-26].

In the low magnetic field regime, evidence for multi-component superconductivity in $UTe_2$ has been observed *via* polar Kerr rotation. The observation of a growing Kerr rotation angle below the superconducting transition temperature, and its trainability with *c*-axis magnetic field, was interpreted as a signature of a time-reversal symmetry breaking (TRSB) multi-component superconducting ground state [27]. Based on this finding, together with the double-peak specific heat anomaly at $T_c$ observed at the time in many $UTe_2$ samples, a two-component superconducting order parameter with the form $B_{2u} + iB_{3u}$ was postulated. Since then, however, a new generation of samples with higher $T_c$ up to 2.1 K were grown. Higher-quality crystals, grown either through chemical vapor transport or molten salt flux, exhibit a single specific heat anomaly at $T_c$ (as well as reduced residual Sommerfeld coefficient $\gamma_0$) [28,29], thereby weakening support for a multi-component superconducting order parameter. Recent polar Kerr effect measurements on samples with single $T_c$ do not show evidence for time-reversal symmetry breaking in the superconducting state [30]. Ultrasound pulsed-echo experiments on samples with both single and double-peak specific heat anomalies also suggest a single-component order parameter, most likely $B_{2u}$ with a single node along the $b$-axis [31]. NMR measurements of a high-quality salt-flux-grown sample of UTe$_2$ observed a large decrease of the Knight shift for $H\|a$, suggesting a single-component order parameter with $A_u$ symmetry, the same as in the spin-triplet superfluid $^3$He isotropic B-phase [32]. In contrast, experiments on the superfluid density probed by penetration depth measurements suggested multi-component superconducting state even in the highest-quality salt-flux-grown samples [33]. The lack of consensus on the nature of the superconducting state in UTe$_2$ may be due to sample variation due to growth methods, as well as experimental conditions, such as application of magnetic field. The question of whether the superconducting order parameter in UTe$_2$ is single- or multi-component, therefore, remains unsettled.

There are general constraints on possible order parameters arising from the symmetry of the crystal structure. UTe$_2$ crystallizes in a body-centered orthorhombic crystal structure within space group *Immm* ($D^{25}_{2h}$). The order parameter at a second order phase transition must belong to an irreducible representation (IR) of the total symmetry group. Therefore, the possible superconducting phases are classified by the crystalline point group symmetry [18]. Basis functions of IRs listed in Table 1 of Ref. [18] (and Table S1 of Supplementary Material) for $D^{25}_{2h}$ are one-dimensional. As a result, if the superconducting order parameter is multicomponent, there is no symmetry argument that guarantees degeneracy between the two superconducting transition temperatures. In contrast, if the superconducting order parameter is single component and odd-parity, then the superconducting state of UTe$_2$ may host point nodes at different high-symmetry points depending on the irreducible representation. A $B_{1u}$ order parameter would display nodes along the c-axis, whereas $B_{2u}$ and $B_{3u}$ would contain nodes along b and a, respectively. Identifying signatures of point or line node(s) in the SC order parameter of UTe$_2$ is an important issue and can narrow down the set of possible ground states in UTe$_2$, as an odd-parity state can either have nodes or be fully gapped according to Blount's theorem [34,35].

Here, we focus on the orbital structure of the order parameter revealed by specific heat measurements in magnetic field up to 12 T applied along the three principal crystal axes. We obtained the residual Sommerfeld coefficient as a function of field, $\gamma_0(H)$, by extrapolating the electronic component of $C/T$ to $T = 0$ K for a given field. Our data show an anisotropic evolution of $\gamma_0(H)$ that indicates that the gap nodes are located close to the a-b plane and remain static in low field (up to roughly 4T). As the field is increased above 4 T our results offer strong evidence that the nodes' positions begin to evolve, at least for the field applied along the a and b axes, with the nodes on the Fermi surface moving away from the direction of magnetic field.

**Experiments details**

Single crystalline samples of UTe$_2$ were prepared using a chemical vapor transport method. The details are described elsewhere [29]. Specific heat data were collected with a quasi-adiabatic technique in an Oxford dilution refrigerator system with a superconducting 12 T magnet. A heater was mounted on one side of a sapphire plate and a high-quality single crystal of UTe$_2$ (with a mass of 5.22 mg and a single $T_c = 2.1$ K) was mounted on the other side with

GE varnish. A ruthenium oxide thick film thermometer was glued directly to the sample. The thermometer was previously calibrated as a function of temperature in magnetic field. A heat pulse is delivered to the heat capacity stage, and the temperature (of the thermometer) is measured as a function of time. Internal equilibration of the stage/sample/thermometer occurred on a much faster time scale than (external) thermal relaxation to the bath, allowing for straightforward determination of the specific heat.

**Results and Discussion**

Figure 1(a-c) shows $C/T$ of UTe$_2$ in various magnetic fields for $H//a$, $H//b$, and $H//c$, respectively. The investigated UTe$_2$ single crystal shows $T_c$ of 2.1 K and a single peak anomaly in $C/T$ in all fields measured [29], allowing a clear determination of $T_c(H)$. The superconducting $H$-$T$ phase diagram for $H||b$ is shown in Fig. 1(d) together with previously reported data. Although the overall $T_c(H)$ is higher compared to previous reports, the $b$-axis magnetic field-dependence of $T_c$ is qualitatively similar in the range of magnetic field used in our measurements [6,36,37]. Fig. 1(e) shows the critical field for all three field orientations and includes the data for $H//a$ from two previous studies [6,38]. $H_{c2}$ is linear in field for $H//a$ and $H//c$, similar to previous reports [6,39]. The extrapolation of $H_{c2}$ to $T = 0$ K for $H//a$ results in a value of roughly 11 T, which is almost two times higher than $H_{c2}$ for lower $T_c$ samples though the increase of $T_c(H=0T)$ is only 30% [6]. These results indicate that the $a$-axis upper critical field is more sensitive to disorder than its counterpart for $H//b$ and that the upper critical field is more sensitive to disorder than $T_c$. Notably, recent theoretical calculations of disordered odd-parity superconducting phases in non-centrosymmetric superconductors find that the critical field is sensitive to disorder whereas the critical temperature is rather robust [40]. The linear tendency of $H_{c2}(T)$ extends down to 10% of $T_c$.

To probe the gap structure of UTe$_2$, we turn to the low-temperature specific heat of UTe$_2$ as a function of temperature in various fields for three different field orientations [Fig. 2(a-f)]. The residual Sommerfeld coefficient $\gamma_0(H)$ for different fields was determined from fits of the data to $C_{fit}(T)/T = \gamma_0(H) + \alpha T^2 + \beta/T^3$, where the $\alpha T^2$ term is expected for an unconventional superconductor with point nodes in the order parameter, and $\beta/T^3$ is the high-temperature tail approximation of a Schottky anomaly.

Figures 2(b,d,f) show the extracted anisotropic field dependence of $\gamma_0$ for field applied along the $b$, $a$, and $c$ axes, respectively. $\gamma_0(H)$ behaves *qualitatively* differently for the three field orientations. However, there are some features that are common for all directions of the magnetic field. We can identify two regimes - the low field and high field regimes, with a boundary at $H_g^i$, where the superscript $i$ stands for $a$, $b$, or $c$ axis, the directions of the applied magnetic field. There are clear kinks in $\gamma_0(H)$ at $H_g^b \approx 4$ T, and $H_g^a \approx 4.5$ T, for field along the $b$ and $a$ axis, respectively, and a more subtle inflection point for $H \parallel c$, with $H_g^c \approx 7$ T.

We examine the low-field region first. For all field directions $\gamma_0(H)$ is sublinear for $H < H_g^i$.

The sublinear behavior of the residual Sommerfeld coefficient in unconventional superconductors is commonly attributed to the presence of the nodes in the order parameter [41-44] and dubbed a Volovik Effect. The enhanced specific heat in magnetic field is due to the Doppler shift $\bar{k} \cdot \bar{v}_s$ of the energy of a nodal quasiparticle with momentum $\bar{k}$ in the presence of a supercurrent with velocity $\bar{v}_s$ associated with magnetic vortices. The energy of the quasiparticles with momentum along the supercurrent is reduced, and the energy of the quasiparticles with opposite momentum is enhanced. This splitting leads to the enhanced density of states around the nodes of the order parameter. An analytical calculation results in a logarithmic field-dependence of $\gamma_0(H) \propto H \log(k_2/H)$, where $k_2$ is a constant [41]. The increase in specific heat was also investigated numerically for states with both lines of nodes and point nodes in the order parameter and for a number of representative directions of the magnetic field [42]. The resulting specific heat was given functionally as $\gamma_0(H) \propto k_1 H^\alpha$. The exponent $\alpha$ was determined to be 0.35 and 0.45 for polar states (nodes along the equator, *a-b* plane) with field parallel and perpendicular to *c*-axis, respectively. For point nodes, $\alpha \approx 0.64$ when the magnetic field is perpendicular to the momentum of the point nodes, and $\alpha \approx 1$ when field is along the nodal direction. In a situation with multiple nodes in the order parameter, nodes that lead to sub-linear field dependence would dominate the response at low field. In general, therefore, the sublinear behavior is expected in the presence of nodes in the superconducting order parameter. The evolution of $\gamma_0(H)$ with field, therefore, provides a powerful tool for investigating the nodal structure of unconventional superconductors (see, e.g. Ref [44]).

The field dependence of $\gamma_0$ for $H < H_g^i$ for all $i$ is well described by $f(H) = \gamma_0^0 + k_1 H^\alpha$ with $\alpha$ close to 0.5, as expected for the order parameter with nodes. Here $\gamma_0^0$ is the residual Sommerfeld coefficient at zero field. Power-law fits as well as logarithmic fits are shown in Figs. 2(b,d,f). The coefficient $k_1$ of the $H^\alpha$ term (the contribution of a particular node to specific heat) depends on the angle between the applied field and the direction of that node. The Doppler shift (and the resulting contribution of the node to the density of states) is greatest when the field is perpendicular to the direction of the node, as the supercurrent $\bar{v}_s$, perpendicular to magnetic field, can then align (anti)parallel with the momentum of the quasiparticles in the node. The coefficients $k_1$ obtained from the fits and shown in Fig. 2(b,d,f) are 12.1, 22.4, and 28.6 for $H\|b$, $H\|a$, and $H\|c$, respectively. The nodes in the low-field regime are therefore closest to the *b* direction and furthest away from the *c* direction. Importantly, the fact that for $H < H_g^{a,b,c}$ the data are well described by single power law fits also indicates that the nodes are static for $H < H_g^i$.

At $H_g^{a,b}$ we observe a sharp upward kink in the data. The power-law fit to the data for $H > H_g^b$ (magenta curve in Fig. 2(b)) results in a negative zero-field intercept (residual zero-field $\gamma_0^0$). A negative zero-field intercept is unphysical, therefore the data for $H > H_g^b$ cannot be described by an order parameter with static nodes as a function of field. Rather, the coefficient $k_1$ acquires field dependence above $H_g^b$. The observed increase above the extrapolated fit to the data in the low field regime can be due to gradual evolution of the position of the nodes with field for $H > H_g^b$. Specifically, the movement of the nodes away from the direction of the magnetic field, becoming more perpendicular to it, will lead to an increase in the contribution of the nodes to the specific heat via Volovik effect.

A similar behavior is observed for $H\|a$ (Fig. 2(d)). There is a sharp upward kink at $H_g^a \approx 4.5$ T. The data curve up quickly to reach the normal value at $H_{c2}$, which is the lowest when $H\|a$. The proximity of $H_{c2}$ to $H_g^a$ may influence the precise functional field dependence of $\gamma_0$ above $H_g^a$. The sharp kink of $\gamma_0(H)$ at $H_g^a$, however, indicates that the nodes in the order parameter begin to move at $H_g^a$, similar to the case for $H//b$. The increase of $\gamma_0$ again suggests that the nodes move away from the direction of the magnetic field toward the *b-c* plane, perpendicular to $H\|a$.

When $H_g^c \approx 7$ T, our data begin to deviate from the $\sqrt{H}$ dependence in the low-field region *via* a mild inflection. Coupled with the expected increase of $\gamma_0$ towards its normal state value at $H_{c2}$, the subtle change at $H_g^c$ indicates that the movement of the nodes for this field direction is much smaller (if any) than for the other two. This is consistent with the observation that the coefficient $k_1$ of the $H^\alpha$ term for $H < H_g^i$ is greatest for $H\|c$, indicating that the nodes are already located close to the a-b plane in the low-field regime.

To summarize, our results in Fig. 2 strongly indicate that the nodes of the superconducting order parameter in UTe$_2$ are located close to or within the *a-b* plane and remain static in the low field regime $H < H_g^{a,b,c}$. Importantly, the nodes are substantially closer to *b* axis than to *a* axis. In the high-field regime, the nodes begin to move away from the direction of the magnetic field. This effect is pronounced for both *H//a* and *H//b*, indicating that the nodes remain largely within the a-*b* plane.

On theoretical grounds, a single-component irreducible representation (IR) order parameter is insufficient to explain even the data in the low field region. A single node in this case must lie along one of the principal crystallographic axes (although see note [45]). However, $\gamma_0(H)$ is expected to be linear for magnetic field applied along that axis, which is not observed experimentally. Recent penetration depth measurements in magnetic field [33] were also interpreted as evidence for the presence of multiple nodes in the superconducting order parameter of UTe$_2$. However, in contrast to our results, the nodes were postulated to be close to both *b*- and *c*-axis. The authors suggested a multicomponent *d*-vector $\boldsymbol{d_{B3u+iAu}} = \boldsymbol{d_{B3u}} + i\boldsymbol{d_{Au}}$ = $(c_1k_xk_yk_z, c_2k_z, c_3k_y) + i(c_4k_x, c_5k_y, c_6k_z)$, where $c_i$ are coefficients of *d*-vectors [33]. The coefficients $c_i$ in multicomponent *d*-vectors can be tuned by the application of magnetic field, thereby changing the locations of the point nodes.

The possible combinations of the two IRs are displayed in Fig. S1 of Supplementary Material (Fig. A·1 in Ref. [46]). We can rule out many of these combinations based on the findings that (1) $k_1^c > k_1^a > k_1^b$, (2) that the nodes move away from the applied magnetic field above $H_g^i$, (3) that nodes lie in the *a-b* plane. These conditions allow us to eliminate possibilities Fig. A·1(a, d, f, i, j, k, l). Next, we make two additional assumptions that (1) in low field the superconducting order has a primary/dominant order parameter, and secondary order parameter is much weaker (smaller magnitude), and (2) that evolution of the relative amplitudes does not move the nodes through the $k_z$ -axis (consult Table II in Ref. [33], also reproduced in Supplementary Material, Table S2). There are then only two possible combinations: $d_{B1u}+id_{B2u}$ with $|d_{B2u}| \gg |d_{B1u}|$, and $d_{B2u}+id_{Au}$ with $|d_{B2u}| \gg |d_{Au}|$ for the low field state, see Fig. S1 in Supplementary Material and Ref. [46]. In both cases, the evolution of the relative magnitudes of the order parameter components keeps the nodes close to within the *a-b* plane (Table S2 in Supplementary Material and Ref. [33]). Fig. 3(b) shows the evolution of the point nodes for $d_{B1u}+id_{B2u}$ with $|d_{B2u}| \gg |d_{B1u}|$ toward $|d_{B2u}| \approx |d_{B1u}|$. The nodes remain in the *a-b* plane and move away from the *b*-axis. Fig. 3 (c) shows the second combination, with two pairs of nodes, where the pair of nodes that lie within the *a-b* plane move away from the *b*-axis towards the *a*-axis. For the $B_{2u}+i\varepsilon A_u$ order parameter the movement of the nodes must conspire to be consistent with the data in the following way: the nodes that lie in the *b-c* plane must move very slowly with increasing field (or perhaps even remain static), while the nodes within the *a-b* plane must move away from the *b*-axis. For the $B_{2u}+i\varepsilon B_{1u}$ order parameter, the two pairs of nodes remain within the *a-b* plane and are consistent with observations as long as the increasing splitting between the nodes leaves them closer to the *b*-axis. On this basis we conclude that $B_{2u}$ is the primary superconducting order parameter in UTe$_2$, which is also suggested by a recent pulsed-echo measurements [31].

**Conclusion**

To summarize, we report the low-temperature specific heat of UTe$_2$ in magnetic field up to 12 T applied along the *a, b,* and *c* axes. Two regimes with distinct responses to the magnetic field are observed. Below the direction-dependent field $H_g^{a,b,c}$, the residual Sommerfeld coefficient $\gamma_0$ displays sublinear field dependence $\gamma_0 = \gamma_0^0 + k_1 H^\alpha$ for all three directions, which points to the presence of static point nodes in the superconducting order parameter. The lowest coefficient $k_1$ is observed for $H\|b$, which indicates that the nodes are the closest to *b*-axis for $H < H_g^b$. Above $H_g^a$ and $H_g^b$, for $H//a$ and $H//b$, respectively, $\gamma_0(H)$ gradually grows above the fits to the data in the low field region, with a more subtle change at $H_g^c$ for $H\|c$. Together with the observation that the largest coefficient $k_1$ occurs for $c$-axis field, the evolution of $\gamma_0(H)$ indicates that the nodes in UTe$_2$ are (1) located close to the $a$-$b$ plane (and closest to $b$) in the low field region and (2) begin to move away from the direction of the applied magnetic field with increasing field for $H > H_g^{a,b,c}$, more evidently for $H \| a,b$. There are two two-component order parameters that can describe our observations: $B_{2u}+i\varepsilon A_u$ with four pairs of nodes and $B_{2u}+i\varepsilon B_{1u}$ with two pairs of nodes. $\varepsilon$ is constant (and small) for $H < H_g^i$, and begins to grow with increasing field for $H > H_g^i$. Based on the discussion above we suggest that the $B_{2u}+iB_{1u}$ is the more likely candidate for the order parameter in UTe$_2$. We emphasize, however, that in either $B_{2u}+i\varepsilon B_{1u}$ or $B_{2u}+i\varepsilon A_u$ scenarios, our measurements indicate that $B_{2u}$ is the primary superconducting order parameter in UTe$_2$. The subdominant order parameter can be stabilized in several ways, including magnetic field (see Supplementary Material), disorder, and sublattice degrees of freedom.

**Acknowledgments**

Work at Los Alamos was performed under the auspices of the US Department of Energy, Office of Science, Division of Materials Science and Engineering. Theoretical work (SZL) was performed, in part, at the Center for Integrated Nanotechnologies, an Office of Science User Facility operated for the U.S. DOE Office of Science, under user proposals #2018BU0010 and #2018BU0083.

**Additional information**

Correspondence and requests for materials should be addressed to RM ([roman@lanl.gov](roman@lanl.gov))

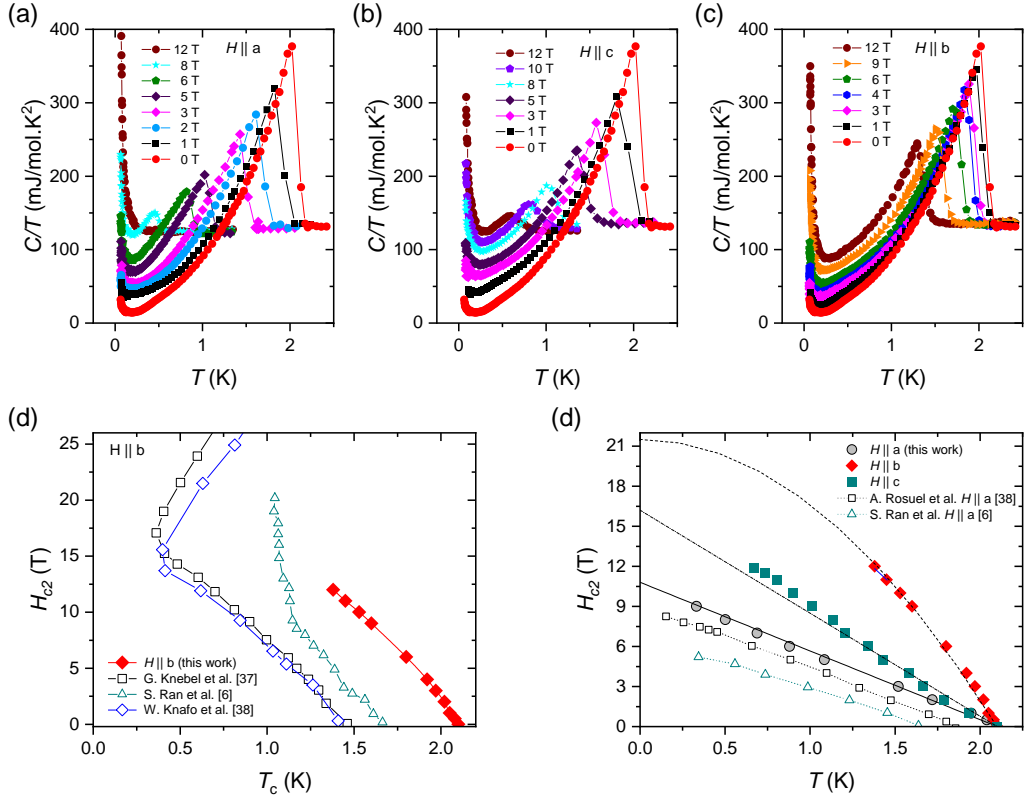

Figure 1. $C/T$ of UTe$_2$ in magnetic field for $H||b$ (a), $H||a$ (b), and $H||c$ (c). Schottky tail at low temperature is enhanced with increasing magnetic field regardless of the direction of the magnetic field. (d) Magnetic-field-temperature phase diagram of UTe$_2$ for $H||b$, plotted with the data reported previously for samples with lower $T_c$. (e) Magnetic-field-temperature phase diagram of UTe$_2$ for $H||a$, $H||b$, and $H||c$, with previously reported data for $H||a$. The solid and dashed lines are fits to the data with $H_{c2}(T) \sim 1-(T/T_c)^m$, with $m=1$ and 2, respectively.

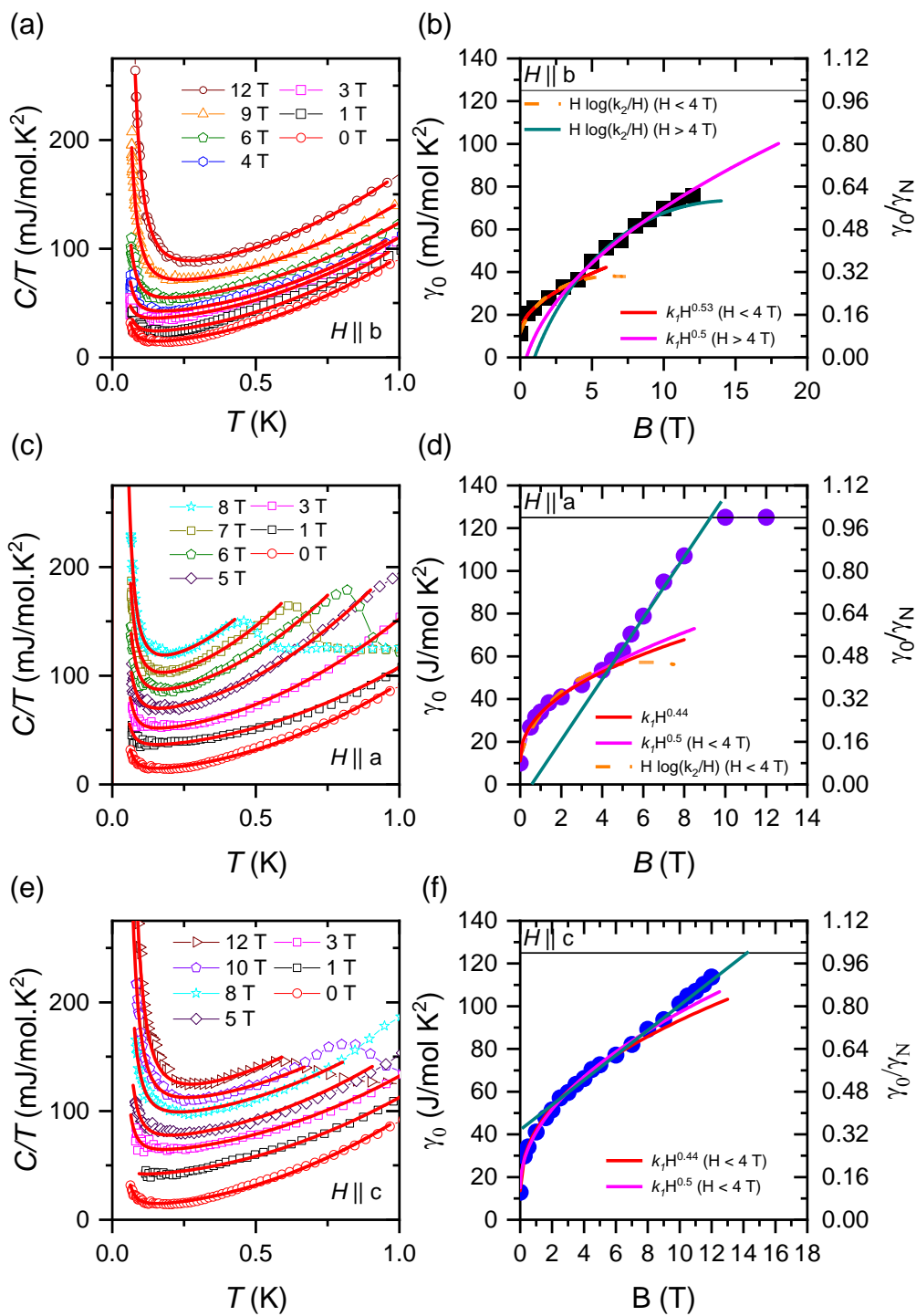

Figure 2. C/T below 1K for three field orientations: (a,b) $H//b$; (c,d) $H//a$; (e,f) $H//c$. In the left panels (a,c,e) $C/T$ is fitted with $C/T = \gamma_0 + \alpha T^2 + \beta T^{-3}$. Right panels (b,d,f) show residual $C/T$ (0 K) = $\gamma_0$ vs magnetic field, with the power law $k_1 H^n$ and logarithmic $H\log(k_2/H)$ fits.

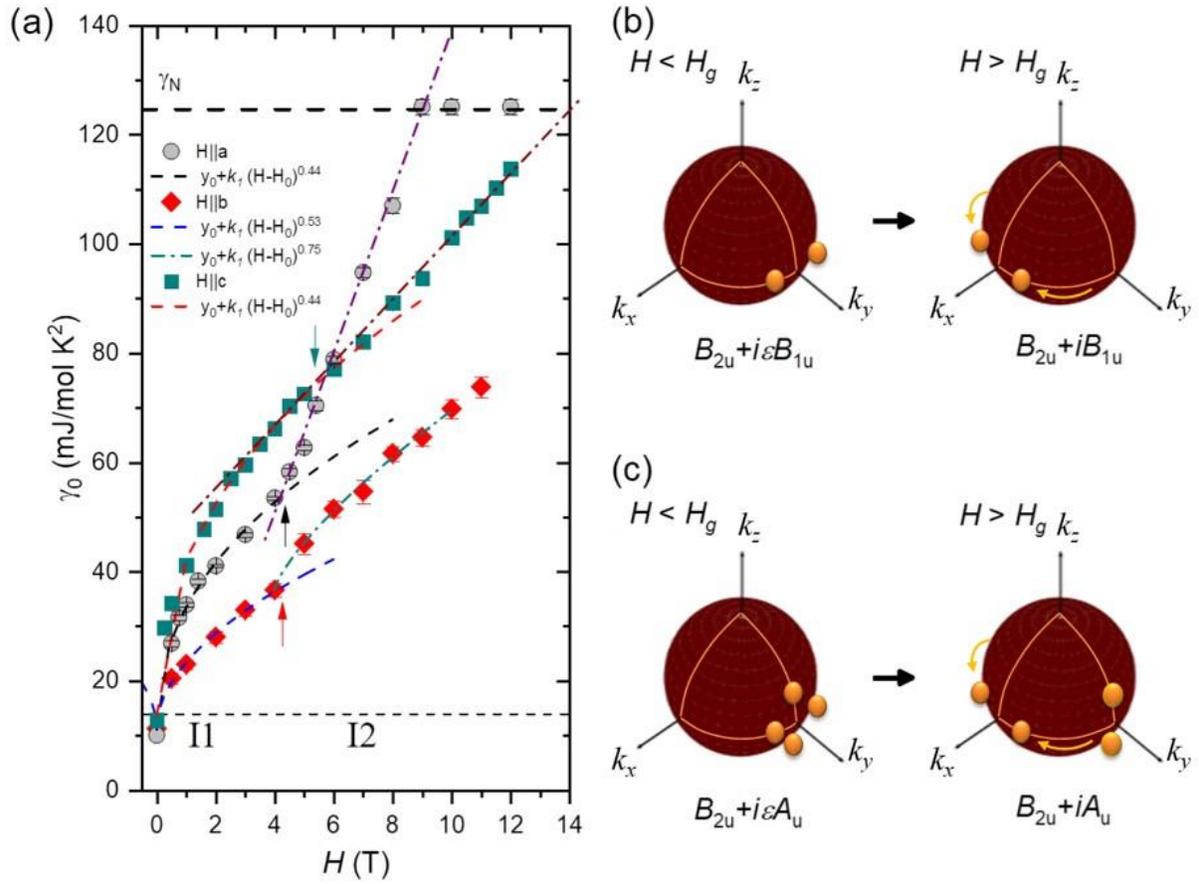

Figure 3. (a) When magnetic field is applied along $H//a$, $H//b$ and $H//c$, residual $C/T = \gamma_0$ vs the magnetic field with $H^n$ fit for $H > H_g^i$ and $H < H_g^i$, respectively. (b, c) Schematics of $B_{2u}+iB_{1u}$ and $B_{2u}+iA_u$ in magnetic field, where $\varepsilon$ is a small constant. The possible locations of the point nodes are marked by the orange spots on a spherical Fermi Surface.